# Designing Workflows for Materials Characterization

Sergei V. Kalinin,[1,a] Maxim Ziatdinov,[2,b] Mahshid Ahmadi,[1] Ayana Ghosh,[2] Kevin Roccapriore,[3] Yongtao Liu,[3] and Rama K. Vasudevan[3,c]

[1] Institute for Advanced Materials and Manufacturing, Department of Materials Science and Engineering, University of Tennessee, Knoxville, TN, 37996, USA

[2] Computational Sciences and Engineering Division, Oak Ridge National Laboratory, Oak Ridge, TN 37830, USA

[3] Center for Nanophase Materials Sciences, Oak Ridge National Laboratory, Oak Ridge, TN, 37830, USA

Experimental science is enabled by the combination of synthesis, imaging, and functional characterization. Synthesis of a new material is typically followed by a set of characterization methods aiming to provide feedback for optimization or discover fundamental mechanisms. However, the sequence of synthesis and characterization methods and their interpretation, or research workflow, has traditionally been driven by human intuition and is highly domain specific. Here we explore concepts of scientific workflows that emerge at the interface between theory, characterization, and imaging. We discuss the criteria by which these workflows can be constructed for special cases of multi-resolution structural imaging and structural and functional characterization. Some considerations for theory-experiment workflows are provided. We further pose that the emergence of user facilities and cloud labs disrupt the classical progression from ideation, orchestration, and execution stages of workflow development and necessitate development of universal frameworks for workflow design, including universal hyper-languages describing laboratory operation, reward functions and their integration between domains, and policy development for workflow optimization. These tools will enable knowledge-based workflow optimization, enable lateral instrumental networks, sequential and parallel orchestration of characterization between dissimilar facilities, and empower distributed research.

---

[a] sergei2@utk.edu
[b] ziatdinovma@ornl.gov
[c] vasudevanrk@ornl.gov

Scientific progress is inherently linked to the development and utilization of progressively more complex methods for synthesis, imaging, and functional characterization of materials, from simple human eye-based examination and macroscopic property measurements to bespoke electron[1] and scanning probe microscopes,[2] scattering facilities,[3-5] and low-temperature quantum measurements.[6-8] These imaging and characterization techniques provide feedback for material synthesis optimization,[7] enable refining of theoretical models,[9] and often lead to serendipitous discoveries.[10, 11] The role of tool development in science is reflected by the renowned quote by Freeman Dyson, one of the leading physicists of the 20th century: "*New directions in science are launched by new tools much more often than by new concepts. The effect of a concept-driven revolution is to explain old things in new ways. The effect of a tool-driven revolution is to discover new things that have to be explained*".[12]

Present day materials discovery, design, and optimization are based on a well-established centuries-old paradigm of serendipitous findings of materials with useful functionalities, and long and time-consuming sequential optimization of compositions and processing conditions towards target functionalities. However, this approach tends to be extremely inefficient in the systems with multiple functionalities that achieve their optimal properties in different parts of multicomponent phase diagrams or synthesis parameter space. One of the materials systems where these limitations are particularly important are the hybrid perovskites for solar cells and other optoelectronic applications.[13-16] Similar challenges emerge for other multicomponent materials and devices including Li-ion batteries,[17] metallurgy,[18] high-entropy alloys,[19] and many functional ceramics and glasses.[20]

The materials characterization workflows typically emerge within specific domain areas, such as the epitaxial thin film growth community,[21] Li-ion batteries,[22] hybrid perovskite solar cells,[16, 23] crystal growth in condensed matter physics[24] or radiation detectors,[25] and many others. For many of these fields, these workflows are inherently familiar to any practitioner in the field and often define it. Once novel or improved characterization tools appear, the workflows adapt to balance the availability of new tools, perceived gains in knowledge, potential for discovery, and costs in terms of time and expenses. This balancing is almost invariably based on intuitive decision making and is constrained by the availability of characterization tools, expected waiting times, and costs.

The new opportunity in the experimental domains is the rise of automated experiments, where the artificial intelligence/machine learning (AI/ML) methods are used both to enable automation to reduce latency (within a domain), as well as guide the discovery workflow. The combination of these two concepts gives rise to the concept of automated laboratories for discovery of new materials for pharmaceutical and biological science and energy applications including solar cells. Despite some early demonstrations, this concept became mainstream only in the last 5 years, as a result of the large-scale efforts by B. Maruyama et al.[26-29], A. Aspuru-Guzik et al.[27, 30, 31], A. Abdolhasani et al.[32-34], as well as (less advertised) efforts in big pharma industries. For the last 3 years, the effort in small scale laboratory-based automated experimentation (AE) via solution synthesis robot resulted in advancement in high throughput and combinatorial studies of hybrid perovskite materials by M. Ahmadi et al.[35-37], C. Brabec et al.[38, 39], etc. However, simple acceleration of materials synthesis by 2-4 orders of magnitude is insufficient compared to the vastness of the composition and processing spaces of multicomponent materials, necessitating development of workflows that will efficiently guide the synthesis and experimental protocols based on the results of previous experiments and general domain knowledge.

As an additional consideration, emergence of new tools gives rise to new opportunities for workflow development. For scanning probe microscopy, examples of this include the development of single molecule unfolding spectroscopy that have opened the pathway to study the kinetics and thermodynamics of single-molecule reactions using benchtop tools,[40, 41] piezoresponse force and electrochemical strain microscopies that have enabled quantitative studies of bias-induced phase transitions and electrochemical reactions at the single defect level,[42-44] and Scanning Tunneling Microscopy for exploring quantum physics[45, 46] and chemical reactions[47] on a single atom level and enabling atomic fabrication.[48] For electron beam methods the examples will include Cryo electron microscopy[49] that enabled mapping of protein structures and hence accelerated drug discovery, and electron diffraction[50] that allowed acquisition diffraction data from very small crystals for crystal structure determination. Examples abound in other fields.

However, until now these developments have been largely *ad hoc*. The workflows have been developed in individual fields[51, 52] and grow and evolve as a result of multi-year community-wide processes. New techniques give rise to fundamentally new scientific opportunities with often rapid growth, but discovery of these opportunities is often a black swan event rather than the result of long-term community-wide planning. Most importantly, in the everyday activity of research groups across academia, government labs, and industry the choice of the measurements and characterization tools is determined by tradition far more then planning or analysis of possible gains and costs.

The workflow development can be subdivided into several elements including ideation, orchestration, and implementation. Traditionally all three elements are human based, and the progression of the scientific career starts with the implementation and progresses to the orchestration and ideation part. The rapid emergence of networks of scientific user facilities and cloud laboratories disrupts this progression, allowing the implementation of workflows via computerized orchestration agents for human and automated equipment. This requires developing systematic ways to design, implement, and build the characterization workflows and determine the gains.[51] Ideally, we want to determine the sequence of measurements in an optimal way, balance the cost of the tools and required characterization times to the knowledge or other gain, and use this to development of characterization workflows.

Here, we describe the scientific workflows, analyze their ideation and optimization for the specific case of the characterization methods, and explore the combined workflows containing characterization and synthesis components. We explore some elements of the workflow building when synthesis, characterization, and theory components are present. Finally, we analyze how these components can be used to enable the next generation of scientific research, including orchestration of the geographically distributed synchronous multimodal characterization workflows, lateral instrument networks, and the emergence of distributed experimental workflows across enterprise level and community level facilities.

**I. Classical scientific workflows: ideation, orchestration, and implementation.**

To illustrate the general concept of a scientific workflow, and some of the general principles of scientific workflow design, here we discuss several examples from areas that the authors are familiar with. However, similar elements and constructs can be identified across multiple other domain areas.

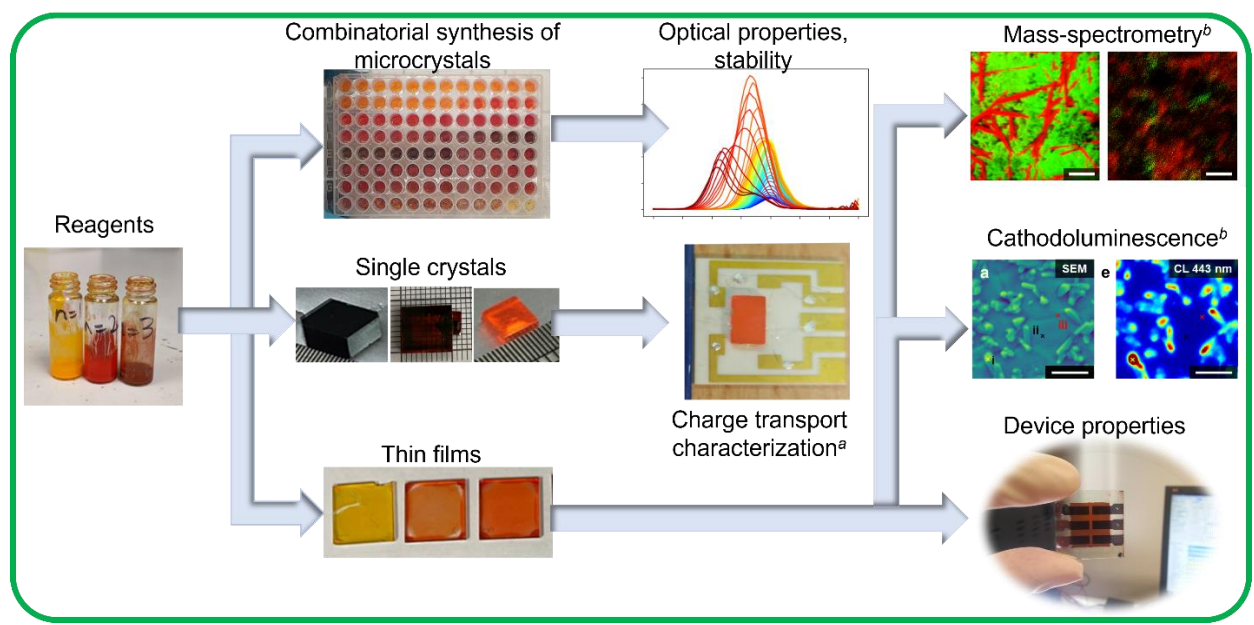

**Figure 1.** Workflow design in the Ahmadi lab focusing on the hybrid perovskite synthesis. Similar reagents are used for the high throughput synthesis and rapid photoluminescent emission assessment of band gap energy, stability and quantum yield, single crystal growth, and thin film deposition. Depending on the final form, the materials can be used to make devices and subsequent physical testing or explored by spatially resolved imaging techniques including cathodoluminescence and chemical imaging. The active devices in turn can be characterized by spatially resolved versions of these methods, providing insight into electronic and ionic dynamics. This forms the backbone of the operational workflow in the lab. The workflow ideation is based on multiple feedback steps where the results of the measurements of properties and functionalities inform the decision making for selecting compositions and processing conditions, stability measurements inform sample selection for film deposition and device fabrication, etc. Most of these decisions are made by human agent and are driven implicitly by combination of intuition on perceived reward, latency of measurement orchestration, and cost. *a*) Reproduced from Ref. [53] with permission from the Royal Society of Chemistry. *b*) Reproduced from Ref. [54] with permission from John Wiley and Sons.

Shown in Figure 1 is the example of the workflow development for the hybrid perovskite synthesis in the Ahmadi lab. Here, shown is the material skeleton of the workflow, meaning the pathway the materials follow during the experiment. Note that workflow can also be defined in terms of the human or instrument time, depending on the specific problem to be solved. The initial point of the workflow development are the initial solutions and antisolvents. These can be used in a pipetting robot to prepare combinatorial libraries spanning the composition space of the hybrid perovskite, antisolvent compositions, etc. The optical band gap energy, materials stability and quantum yield within the libraries are in turn accessed by the time-evolution of photoluminescent and UV-Vis absorption spectroscopies. Based on the results, the specific composition can be chosen for imaging studies with high spatial resolution. Similar solutions can be used for the single crystal growth. The crystals can be fabricated into the devices for radiation sensors, explored via photo-Hall effect spectroscopy, neutron scattering or Mossbauer spectroscopy. The crystal-based devices can in turn be visualized in operando via scanning probe microscopy or time-of-flight mass-spectrometry (ToF SIMS). Finally, the same initial reagents can be used to deposit thin films. The films in turn can be further fabricated into the device structures that can undergo physical testing. The microstructure of the films can be explored using cathodoluminescent (CL) imaging, ToF-SIMs, and scanning probe microscopy. Finally, devices can be characterized in operando using same techniques as crystals.

The characteristic aspect of this workflow is the progression from rapid, low cost, and high-throughput methods that provide limited and low-fidelity information on specific functionalities, to the expensive slow characterization methods for end materials or devices. In this process, the orchestration of the workflow includes multiple decision making and feedback steps on what composition to choose for complex characterizations. Note that the decision making is often non-linear. For example, the results of the photoluminescence (PL) screening can be used for the composition selection for film deposition, and the CL and ToF SIMS imaging of films will be used for composition selection for the robotic synthesis or the material selection for the initial endmembers or solvents. This decision making is further informed by the general information available to the researchers and adjusted as a result of the interaction with the scientific community via publications, conferences, social networks, interaction with large language model optimized for generating hypotheses (a scientific version of ChatGPT), and private communications.

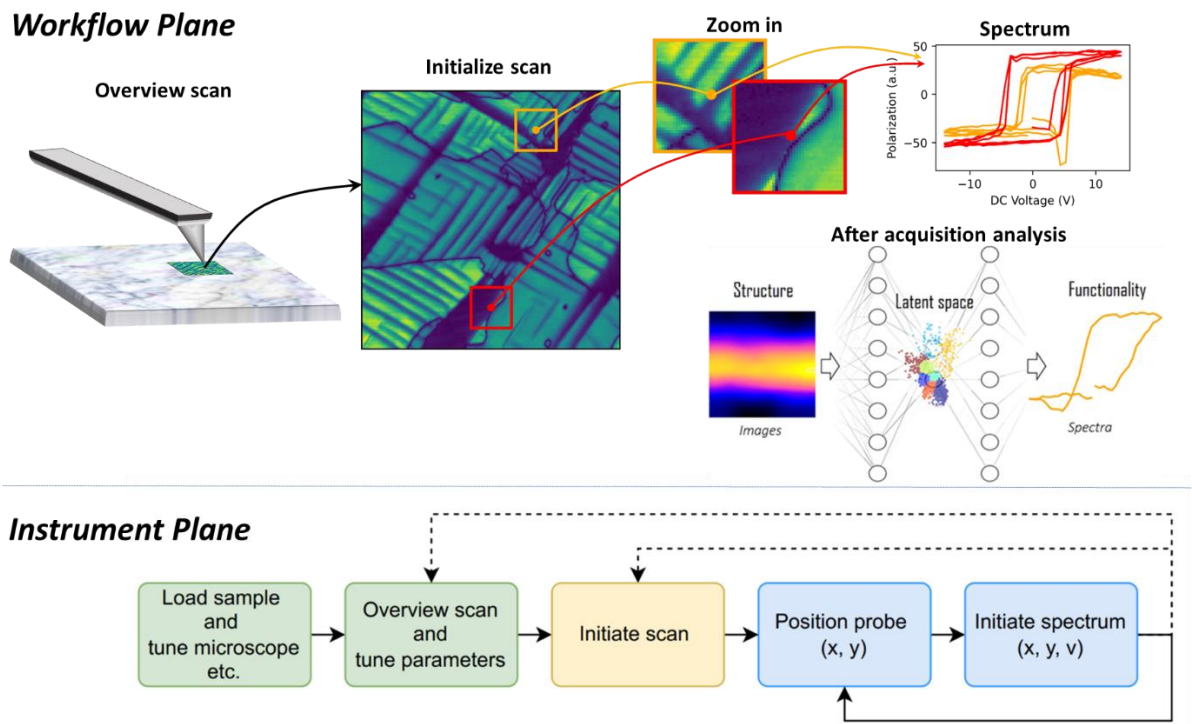

**Figure 2**. Workflow development in imaging. Here, the decision making includes the selection of specific regions for high-resolution studies and subsequently for the spectroscopic probing. This process is often iterative and includes multiple overview scans, zoom-in and zoom-out stages, and human-driven spectrum acquisition and hyperspectral imaging. Note that for many instruments, the workflow will also include the repetitive tuning of instrument parameters for maximizing instrument performance. In this case, the workflow skeleton represents that sequence of operations performed by the microscope, and workflow orchestration is effectively a stochastic optimization process. The nature of the reward function and values of individual steps are dependent on human operator.

The second example shown in Figure 2 is workflow development in imaging.[51] Here, the natural backbone of the process is the series of operations performed by the microscope based on the input from human operator and internal feedbacks. The initial state of the system comprises the chosen sample and operator knowledge. The instrument operation includes tuning of the imaging conditions and overview scanning. Based on observations, the operator makes a decision to zoom-in on specific regions to explore these in detail, and potentially zoom-out and zoom in on a different region and perform spectroscopic measurements. We note that for the unknown systems (for which there are no prior measurements or sufficient literature data) the selection of zoom-in regions is usually based on the educated guess of an operator and what he/she considers to "look interesting." Depending on the type of the measurement and instrument configuration, the operator can perform spectroscopic measurements on the grid (hyperspectral imaging). The process can also include the additional microscope tuning steps, with the human operator selecting the times to introduce them based on the operator assessment of the observed data.

There are three pertinent elements of the described process. The first is a clearly defined hyper-language that describes the human-initiated high-level operations. This language can differ between domain areas, but for most human-operated tools the set of elementary commands are similar. The second is the workflow orchestration based on the responses generated during the experiment and evaluated by human operator, including the decisions of what region to select for scanning and spectroscopy, and when to tune the instrument. Note that instrument automation makes some of these operations automatic, and often prompts the operator to perform them. The third and key element is that of the (scalar or multi-objective) reward function. The operator interested in mechanical properties of the material will be interested in different objects than the person exploring the emergence of ultra-high electromechanical responses. Similarly, scientists interested in ferroelastic phenomena will choose different object for study than those interested in flexoelectric phenomena at ferroelectric domain walls. This reward function in turn determines the value of the individual steps for the operator, and in this fashion guides the workflow ideation. Finally, note that the scientific workflows and rewards have a clear hierarchical character, obvious given that example in Figure 2 represents any of the imaging techniques that are used as a part of workflow in Figure 1.

The brief examination of these workflows illustrates several common elements. The first is the emergence of the funnel character. All samples are explored using easy characterization methods such as optical microscopy (and sometimes just human eye) or low-resolution overview scans, and this information is used to select locations for progressively more complex or expensive methods for smaller number of samples or locations. The second and less obvious component is the perceived reward of the experiment. In some cases, this is purely curiosity-based selection. In others, this is targeted exploration of specific aspects of microstructure or materials behavior. In this latter case, the exploration pathway is driven by the specific interest of the experimentalist. The third implicit component is reliance on prior knowledge in selection of objects of interest, interpretation, and establishing the reward. Similarly, the discoveries can update the knowledge base. While slow for characterization experiments, this strongly affects materials synthesis workflows shown in Figure 1.

The two examples above illustrate the workflow concept, organization of the workflows based on materials or instrument time, and the complex hierarchical connections between the reward functions of the individual elements and values of the specific steps and operations. Below, we explore the principles based on which, these workflows can be designed and optimized – here, for the specific case of characterization workflows as shown in Figure 2. However, we emphasize that these workflows are in turn defined and have value only in the context of general synthesis and characterization workflows shown in Figure 1. In discussing these workflows, we separate the components of imaging, spectroscopy, and theory. Here, imaging is referred to the process of the acquisition of spatially resolved structural and functional information. Spectroscopy refers to the process of the detailed measurement in a specific location that is assumed to provide desired information on materials functionality. Note that the process is hierarchical, in a sense that imaging can be performed via the acquisition of spectra on the sample grid (hyperspectral imaging) if the spectroscopy is non-destructive. Due to the difference in cost and latencies, this gives rise to such standard tasks as pan-sharpening.[55] Finally, the third component is theoretical analysis, generally referring to the derivation of the insights from observations and using these to modify the way workflow is ideated. First, we discuss the principles of workflow development for pure imaging and spectroscopy scenarios and compare it to the (well explored) ideation of theoretical workflows.

**II. Single element workflows**

We define the single element multiresolution workflow as those emerging within a single type of characterization or theory. In fact, the concept of multiscale workflows is well developed in the theory domain. Over the years, the development and implementation of multiscale modeling[56-63] workflows have paved their way into multiple disciplines ranging from solid mechanics, fluid mechanics, materials science, physics, mathematics, biological to chemistry. Parallel computing has only made it more feasible to solve more accurate and precise algorithmic formulations which is needed for these workflows. In addition to being useful in academic research, such modeling capabilities have been adapted in industry for its numerous advantages such as cost-effective physics-based product design, assessment of product quality and performance. The primary requirement for any multiscale workflow is to lay out strategies to bridge between different length scales, going from atoms to automotives. We note that it differs from the conventional point of view being followed in various disciplines where the focus remains on solving a particular challenge with sole consideration of the pertinent length scale and associated latencies. Two of the most common approaches as followed in the multiscale paradigm are concurrent[64, 65] and hierarchical[66] in nature. Methods to bridge between length scales vary between the two. In concurrent techniques, the bridging methods depend on numerical solutions, whereas hierarchical approaches rely on performing independent numerical simulations at different length scales followed by identifying relations between parameters relevant to integrate or reconstruct materials behavior at the corresponding higher layer in the ladder of length scales. The hierarchical approach is top-down in nature. Understanding microstructure-property relationships inferred from the interplay of internal state variables existent in various scales with thermodynamic constraints is a good illustration of such an approach. One example where concurrent and hierarchical schemes are practiced together[59, 67-69] is in connecting atomic scales simulations with electronics principles scale simulations. Density functional theory (DFT) simulations performed on metals are coupled with embedded atom method (EAM) potentials within molecular dynamics (MD) environment to model edge dislocations, with subsequent studies performed with quantum models at disparate length scales. Here, the many body interactions are evaluated within the semi-empirical formalisms of the potentials such that the results from the electronic structure theory computations become useful to reproduce physical properties of many metals, defects, and impurities. In addition, it is also possible to conduct such multiscale studies on-the-fly where the classical potential adapts to the local environment via dynamic force matching. Several machine learning frameworks have also proven to be useful to develop these potentials[70-76] within an interactive suit bridging between the atomic, coarse-grained descriptions with promises of connecting to the continuum theories.

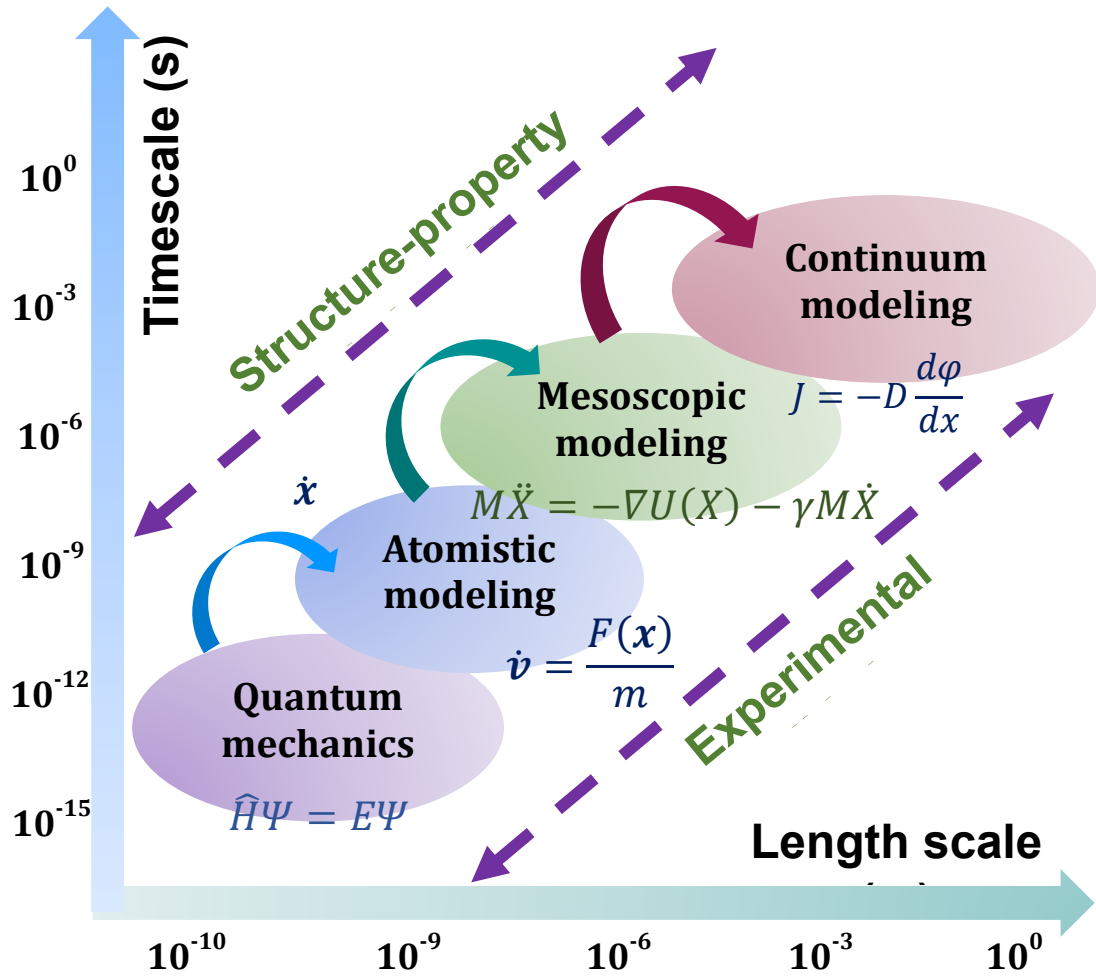

**Figure 3.** Multiscale workflows in theory

Machine learning methods have significantly accelerated the development of multiscale workflows for theory. Multiple approaches to address it have been used, ranging from rare event samplings in molecular dynamics (MD) simulations to machine learning based information compression schemes[77, 78] Particularly over the last five years, a number of machine learning approaches based on variational autoencoders,[79] generative adversarial networks, and diffusion models have been suggested to bridge length and time scales in simulations, establish statistically significant descriptors such as order parameters, and determine their constitutive relations. It should be noted that many of these methods have also been shown for the information compression in experimental data such as electron and scanning probe microscopy,[80-82] albeit with the additional requirements to account for the out of distribution shifts due to changes in imaging conditions that are typically absent in modelling.

## II.A. Data driven multiresolution discovery

The nature of the multiscale problem in experiment is opposite in the sense of information flow. Multiscale characterization in experiment requires solution of the inverse problem, namely the optimal hierarchical experiment design. For example, given the results of macroscopic measurements or low-resolution imaging, we seek to identify the potential object of interest to be explored via high-resolution imaging probes in such way as to gain the maximal insight into the nature and origins of observed macroscopic behaviors.

To set this problem in a more mathematical basis, we consider it as one of optimizing a sequence of actions to maximize a cumulative reward function.[83] The reward function can be defined in multiple contexts from pure optimal structure discovery (we aim to characterize the structure as detailed as possible via reduced multiscale representations) to discovery based on prior knowledge (we know what microstructural objects we are interested in). In case of information gain, one can simply seek to minimize uncertainty arising from models at different length scales, by finding the sequence of measurements that can best reduce the overall uncertainty across length scales. Alternatively, a measurement sequence can be found that minimizes the uncertainty at one desired length scale. To construct a reward function for this can be as simple as taking the negative of the sum of the uncertainty. In a reinforcement learning framework, the actions consist of both the length scale to explore next in the workflow, as well as parameters within the specific experiment. For simplicity we will ignore the latter component and consider that the only action is to choose the appropriate length scale. The task is to find a policy that will determine how best to select actions to maximize the cumulative rewards. Additionally, it is known that measurements at one length scale can be highly informative of measurements at other length scales.

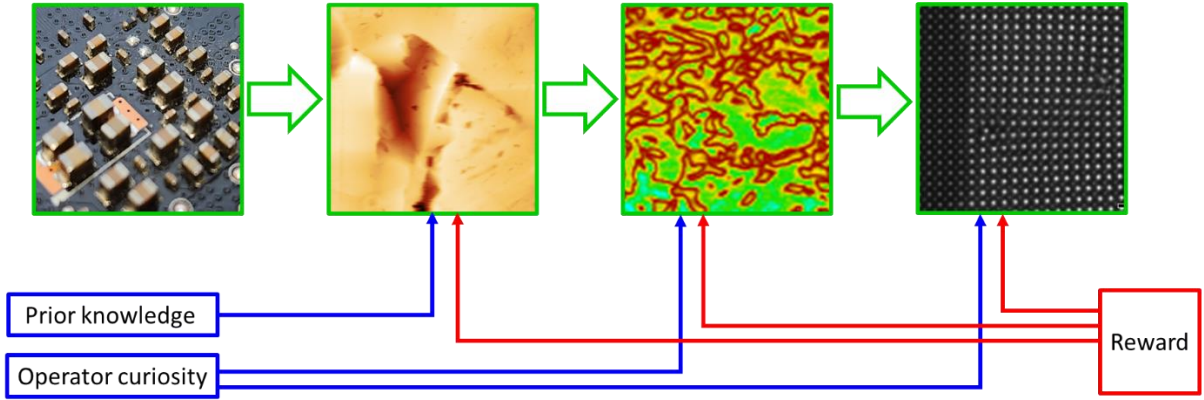

**Figure 4.** The workflow for structural imaging. Here, the imaging studies are performed increasing the resolution of the microscope (and changing the imaging system). The decision making along the workflow includes selection of regions for detailed studies at progressively higher level of details. The decision making along the workflow can be purely data driven (green lines), or incorporate decisions made with prior knowledge and informed by perceived reward.

For this scenario, the knowledge discovery process can be represented via a probabilistic machine learning (PML) framework based, for example, on Gaussian process, Bayesian neural network, or deep kernel learning, with the idea that the data can be mapped from one length scale to another. Assume that we have training data X captured at the $n^{th}$ level, $X^n$. We can generate predictions on the test data $X_*^n$,

$$X_*^n \to \bar{f}_*^n, \mathbb{V}(X_*^n) \qquad [1]$$

As an example, consider that we take a few optical images at certain positions ($X^n$), and then want to predict the image at unseen locations ($X_*^n$). The predicted images are $\bar{f}_*^n$, and the predictive uncertainty is given by , $\mathbb{V}(X_*^n)$. But we can also setup models that attempt to predict at the next (e.g., lower) length scale. By feeding in the test data $X_*^1$, we can train a PML model to predict the function value at the next level, i.e., $X_*^1 \xrightarrow{PML^{(1)}} \bar{f}_{X_*^1}^2, \mathbb{V}^2(X_*^1)$ . In general, we have

$$X_*^m \xrightarrow{PML^{(m)}} \bar{f}_{X_*^m}^{m+1}, \mathbb{V}^{m+1}(X_*^m) \qquad [2]$$

The task is to determine which measurement level *m* (out of the *n* available levels) to choose that will minimize the uncertainty in $\mathbb{V}(\cdot)$. however, as mentioned before, we need to know how to set the reward. This means we need to know which $\mathbb{V}_m$ or combinations thereof to use. Minimizing this uncertainty will be the objective of the policy. After fixing the reward function, the workflow reduces to a simple reinforcement learning (RL) environment in which the action is to decide which measurement (level) to capture, which will then enable the appropriate PML[(m)] model to be updated. The state fed back to the agent will then be the set of predicted means, i.e. $\bar{f}_{X_*}^k$ for $1 \leq k \leq n$. In the case where we don't know which $\mathbb{V}_m$ to use, this can be reformulated as an RL problem, with the difference being that the Action now selects not only the measurement level, but also, which uncertainty map level is used. Alternatively, one may also consider some type of superpositions of $\mathbb{V}_m(\cdot)$ from different levels. Additionally, given that different experiments will likely greatly differ in their actual cost per data point measured, this can also be factored into the reward function: penalties can be applied to actions or action sequences that use 'expensive' characterization tools, for instance. Regardless, the optimization is straightforward once the reward is defined, similar to our recent work with hypothesis learning.[84, 85]

As a practical example, consider that we might have optical images, SEM images with some chemical maps (e.g., from energy dispersive spectroscopy, EDS), and some scanning tunneling microscopy images of a material system, such as a substrate with 2D flakes of varying chemical composition. This defines three levels, and the question is which set of experiments will be most informative. In this case, optical microscopy of similar looking flakes is not likely to be highly correlated with local electronic structure as imaged by STM, if these flakes are different in chemical composition and defect concentrations. On the other hand, there will be a significant correlation with the EDS data on the same flakes. As such, after initial correlations between STM, optical and SEM/EDS data is found, minimizing total uncertainty may hinge more on performing a few experiments with SEM/EDS and confirming them with STM. Our problem is in determining exactly which of these measurements are most informative, and this can be done in principle, via the PML framework described above.

## II.B. Beyond data driven discovery

The recent advances in imaging and characterization tools including electron and scanning probe microscopy and associated spectroscopies, atom probe tomography, focused X-Ray scattering, nanoindentation, optical microscopy and a gamut of electrical, mechanical, and magnetic testing methods span the multitude of length scales and functionalities. However, the gamut of available techniques is belied by the dearth of the systematic workflows that allow exploring the behaviors of interest in a systematic and unbiased way. It is by now common to complement the macroscopic probing of the piezoelectric, catalytic, and electric properties by the STEM or atomic probe tomography studies of atomic structures, or correlate the photovoltaic performance of polycrystalline solar cells by the nanometer resolution cathodoluminescence and chemical imaging maps.

However, it is seldom when we can say with the certainty that it is the specific type of microstructural elements most strongly associated with the functionalities of interest. In the cases where such relationship were defined, as for the role of step edges for catalysis, dislocation theory of plasticity, or role of the domain wall dynamics on giant electromechanical responses in piezoelectric ceramics, these discoveries required community-wide effort involving multiple experiment and theory development cycles. At the same time, it is these insights that are most relevant towards understanding the underpinning mechanisms and particularly establishing pathways for materials optimization and discovery. Only by understanding the fundamental origins of structure-property relationship can the strategies for the improving materials performance be formulated and tested.

For example, given the polycrystalline ferroelectric material, we may seek to understand the origins of the high electromechanical response or resistive switching. Given the hybrid perovskite solar cell, we want to understand which microstructural elements are responsible for chemical stability, current-voltage hysteresis, or the open circuit voltage (OCV) losses compared to the ideal values. The complexity of such analysis stems from the fact that it may be different defect populations that are ultimately responsible for these behaviors, and hence selection of the regions of interest for the detailed studies depends on the specific goals. In this sense, this problem is poorly defined – we seek to preferentially explore via detailed high-resolution studies objects whose identity we do not know. Hence, we use hypotheses formed based on the prior body of knowledge to guide the exploration process, while maintaining the need for serendipitous discoveries.

**III. Multiresolution structure-property characterization workflows**

It is well recognized that understanding structure-property relationships in materials requires exploring properties and functionalities on all length scales, from the atomic scale structures to mesoscale to the global properties of the material or device. Multiresolution analysis workflows are used when, for each characterization method, we have access to image-spectroscopy pairs, and the spectroscopic data provides information that is predictive of the functionality of interest. Here, we define as the image the high spatial resolution/low information density imaging including structural STEM image, SPM images, etc. The spectroscopy refers to the information rich local measurements that are associated with larger measurement times or lead to the irreversible changes in materials structure, and therefore can be performed only in a limited number of locations. However, we implicitly assume that the spectroscopic measurements are correlated with the macroscopic functionality of interest.

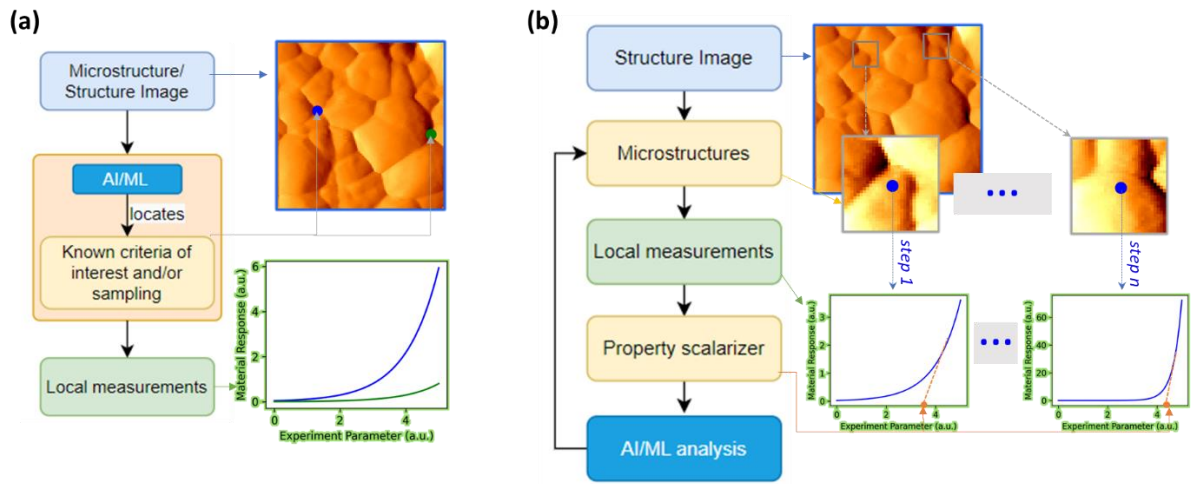

**Figure 5.** Single step workflows based on (a) known criteria of interest and (b) active exploration, also known as (a) forward and (b) inverse experiments. In the case of (a), the object of interest are known a priori and the role of AI algorithm is to identify them and perform specific measurements. In case (b), the scientist is interested in certain aspects of the functional properties defined via scalarizer, and the AI algorithms learns what microstructural elements maximize or minimize them.

**III.A. Single step direct workflow**

Here, first we illustrate these concepts for the single structure-property mapping step. In this case, we can define *forward* and *inverse* experiment. The forward experiment relies on the a priori defined objects of interest that can be recognized in real time, e.g. using deep convolutional networks. Here, the emergence of the ensemble and iterative training methods allowed to partially address the inevitable out-of-distribution effects (i.e. capability of the trained network recognize object of interest if microscope parameters have changed). Recently, a deep residual learning framework with holistically-nested edge detection (ResHedNet) was ensembled to minimize the out-of-distribution drift effects in real-time SPM measurement.[86] The ensembled ResHedNet was implemented on an operating SPM, where it converted the real-time SPM data stream to segmented objects of interest, e.g. ferroelastic domain wall or polycrystal grain boundary images. Then, a pre-defined workflow used the coordinates of the discovered objects for spectroscopic measurements. In doing so, the approach allowed a thorough of interested objects of interest (virtually all locations at objects of interest) in an automated manner, in contrast, traditional manual operation only allows us to investigate a limited number of locations at domain walls. Using this approach, alternating high- and low- polarization dynamic ferroelastic domain walls in a $PbTiO_3$ thin film was observed[87] and the behavior of grain boundary junction points in metal halide perovskites was discovered.[88]

### III.B. Single step inverse workflow

In the inverse experiment, the operator defines the characteristics that make spectrum "interesting", e.g. intensity of a specific feature, specific aspect of spectrum shape, or even maximal variability of spectra within the image. In other words, each collected spectrum can be associated with a single number defining how interesting it is, either in absolute sense or as compared to previously acquired spectrum. The deep kernel learning (DKL) algorithm learns what elements of the materials structure maximize this reward and guides the exploration of materials surface accordingly. This DKL algorithm was recently implemented in SPM to investigate the relationship between ferroelectric domain structure and polarization dynamics,[89] and in STEM to explore bulk and edge plasmonic modes. As show in Figure 2, the DKL exploration process identifies the domain walls as objects of interest and the DKL predictions indicate the high polarization dynamic of 180º domain walls. Although these are expected by ferroelectric experts, the DKL itself did not have any prior physical knowledge, and learned all that information during the experiment.

### III.C. Multiple step workflows

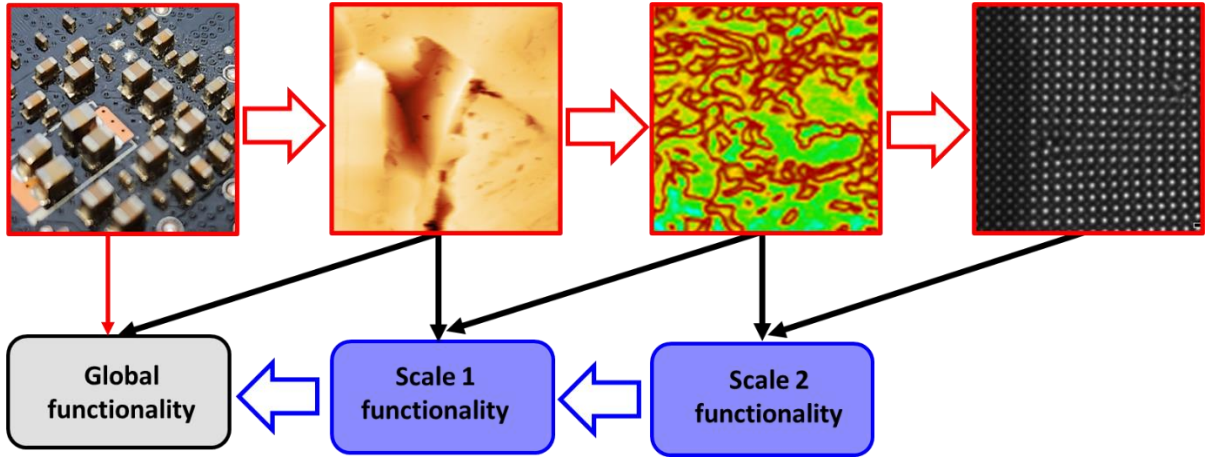

**Figure 6.** Multiple level imaging and characterization workflow. Here, we assume that the material functionality of interest probed on the macroscopic level are controlled by the hierarchy of structural elements from macroscopic to atomic scale, and that the available functional probing methods is representative of the materials functionality. For example, on the mesoscale this can be realized using the microelectrodes and the SPM-based IV measurements.

The multiple step imaging and characterization workflows can be represented as the direct extension of single step workflows as shown in Figure 6. Here, the structural imaging at the low resolution yields the library of possible microstructural elements. We assume that the microscopic measurements, e.g. using the micropatterned contact arrays and current-voltage (IV) measurements via SPM, are representative of the macroscopic properties of the systems (even though the exact physical mechanisms can be different due to changes in contact conditions, confinement effects, etc.). These elements can be sampled in statistically balanced way, e.g. based on the distributions in the latent space of the system, to give the initial information on the structure-property relationships in the system. With these, the process can be iterated balancing the structural learning and learning structure-property relationships on multiple length scales.

For the cases when the spectroscopic measurements provide information that is a proxy to the macroscopic functionalities but does not directly reproduce it, the definition of the workflows become more complex and requires incorporation of the multiresolution multifidelity measurements.

**IV. Some considerations on theory-experiment-characterization workflows**

The particularly interesting problems emerge at the interface between the theory and experiment. The conflux of parallel computing, experimental capabilities along with theoretical simulations is necessary to develop and implement end-to-end theory-experiment-characterization workflows. It is safe to say that within materials science challenges spanning over various applications, the primary goal is to leverage existing structure-property relationships[90] to propel both design and discovery. These relations could be formulated at the atomistic level where electronic structure of systems plays the key role to determine the energetics and stability. For mesoscale to continuum scale, we tend to map the evolutions of microstructures with physical properties such as plasticity, damage, or failure. The standalone theoretical investigations are well capable of elucidating the fundamental mechanisms responsible for materials characteristics. However, the ultimate validations of such proposed mechanisms always rely on experimental observables. Hence, to fully realize the potential of workflows[91-98] bridging instruments and theory, we must move towards theory-assisted experiments from the standard perspective of matching final outcomes from experiments and simulations. It becomes important to consider how causal structure-property relations may hold true at different length scales while establishing connections between experimentally controllable parameters with theoretical variables. Here, we consider this interaction only in the more limited context of imaging and characterization methods, but even in this case the complexity of possible interactions is immense.

Here, we assume that we have access to microstructure (M) at a single length scale, global property measurements (G), local property measurement (L) and theoretical model (T). We further assume that M and G are available from the beginning of the experiment and are not updated. Comparatively, T is available in the form of analytical or numerical model and can contain partially unknown parameters that can be updated during the experiment, and L can be performed sequentially within known M. With this, we can define the static learning problem, meaning establishing the relationships between G, M, and T after the measurements, and active learning problems, meaning the workflow design for active experiments within M. Here, we will use arrow → to define establishing a relationship given full data, and ► to define active learning workflow.

**IV.A. Static problems**

With these building blocks, we can combinatorically define three static analysis problems. Here, MT→G is a direct calculation problem. In this case, we assume that the microstructure is known and the theory is correct, and aim to calculate the global properties of the system. The example of such approach will be the finite element calculation of the mechanical properties of the composite material, estimation of the effective transport properties of the microstructure mapped via the X-Ray tomography, etc.

The inverse problem will be MG→T. Here, given known microstructure and global properties, we aim to refine theory that governs structure-property relationships in the system. Finally, GT→M defines the design problem. Given the property of interest and theory, we aim to design microstructure satisfying the given properties. These problems are static in nature, i.e., we do not have iterative/active learning component. Of course, these problems become active when a part of the synthesis or manufacturing workflow.

**IV.B. Active learning problems:**

With these simple examples, we can define several classes of active learning problems. As mentioned above, defining active learning workflows necessitates introduction of the reward, R, defining the discovery target. For example, M(R)►L is the direct imaging problem, where we choose locations for sampling local microstructure L given what is interesting about M. In our experience for human driven exploration the reward function often changes during the experiment. For example, initial measurements are done with the target of microscope performance optimization, proceed to get the overview image, explore the most statistically prevalent regions withing the image, and then proceed to explore the regions that are believed to be interesting based on the prior knowledge and beliefs. For example, it is natural to target dislocations for mechanical properties, or ferroelastic domain walls for understanding of the origins of the ultra-high electromechanical responses.

Comparatively, the L(R)►M is the DKL problem, where we aim to discover locations for L given what is interesting about L and perceived reward R. We can also envision more complex workflows, for example ML►T, meaning how to learn the theoretical model in an efficient manner given microstructure and local measurements. The more complex scenarios include M G(R)►L , M T(R) ► L, and M GT(R) ► L, meaning that we aim to discover locations for L given what we know about microstructure and theory and perceived benefit for global properties, theory, or both. As defined earlier, the reward can be optimization of some characteristic, or simply, information gain as represented by reduction in uncertainty of DKL (or other relevant) surrogate models.

For multilevel problems, actively learning theory at the same time will induce additional nonstationary characteristics to the objective, further increasing the complexity of the problem. In fact, even the simple case of the co-navigation of theoretical and experimental domain requires detailed analysis.

## V. Implications

Finally, we analyze how the design and optimization of synthesis and characterization workflows can affect the structural organization of the research process. Specifically, we consider both the driving forces such as increase of the throughput and costs of many imaging and characterization tools, and emergence of the cloud infrastructure and edge computing that allow information processing and feedback from the cloud.

### V.A. User facilities and lateral instrumental networks

From the historical perspective, the first big changes in the scientific workflow in condensed matter physics, materials science, and biology were brought about by the emergence of the large tools such as synchrotrons and nuclear reactors. At that time, the concept of user facility included developing the instruments, instrument scientists operating them, and user scientists that physically visit for day-week long experiments on their specific materials has become a norm.

The second big change has been emergence of the user facilities, as exemplified by the Department of Energy Nanoscale Science Research Centers, and user facilities at universities. This trend is associated with the emergence of the microfabrication labs as a part of exploratory research, and rapid growth of throughput and costs of characterization tools such as electron microscopes, scanning probes, and chemical imaging. Correspondingly, integration of these tools within the same geographically localized facilities that maintain the instrument, offer the sample preparation facilities often shared between multiple instruments, maintain the highly-trained scientists capable of operating and using them greatly increases the efficiency of use.

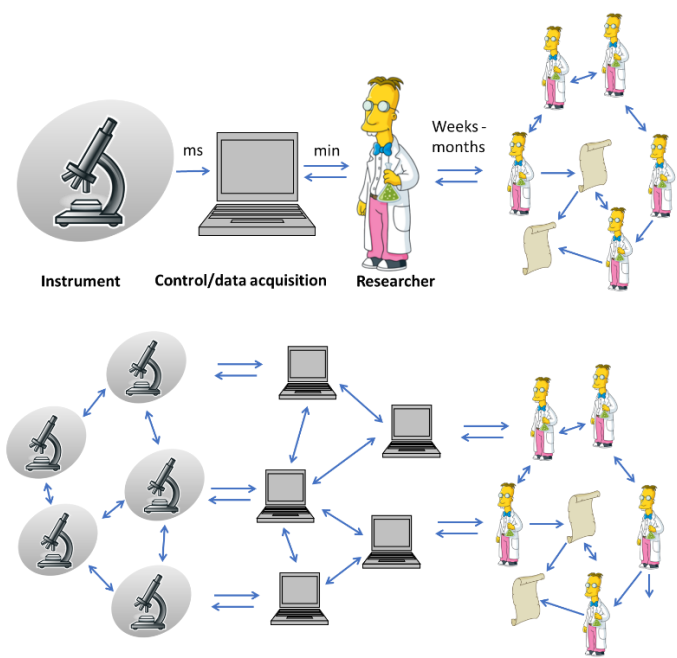

**Figure 7.** Transition from single tool to lateral instrumental networks enabled by the cloud technologies and the data analysis ecosystem.

However, despite the drastic changes on the operational side, the mode of use of the instruments in user facilities and individual labs have been remaining largely the same. In all cases, the scientist run the instrument manually, generating the large volumes of data during the experiment. The data is typically analyzed after the experiment, in the process that often takes weeks and months, and is subsequently shared with the scientific community via publications. The latter process is typically associated with extremely large latencies of the order of months and often years, hindering the research process. Correspondingly, until 10-15 years ago scientific conferences were the primary means for rapid information sharing. Over the last decade, rapid growth of popularity of the preprint servers such as arxiv, chemrxiv, and biorxiv, as well as social networks greatly accelerated information sharing. Similarly, code and data sharing via platforms such as GitHub, Zenodo, and Google Colab is rapidly becoming a part of scientific culture in many domain areas.

The rapid increase of ease of use of cloud technologies suggests that the field is now poised to the next transition, where the operation of the tools is largely remote and the information is either directly streamed to the cloud storage, streamed after the acquisition based on the upload speed constraints, or following the point-of-generation compression. In addition to obvious advantages in terms of data accessibility and security, this allows the formation of the lateral instrument networks as illustrated in Figure 7. Here, multiple instruments of the same kind store the data within a community-accessible cloud space. The latter also supports the computational capabilities and code ecosystems that allow data analysis, and in turn can be further harnessed for the decision making and automation of workflow on individual instruments, significantly accelerating the scientific process. It should be noted that very likely data permissions will be dependent on the study; most likely, data will not be shared universally without an embargo period' or something of the sort.

**V.B. Sequential and parallel experiment orchestration**

The development of cloud connectivity for characterization tools establishes a set of novel opportunities for experimental workflows across multiple facilities. The particularly important case for the parallel experiment orchestration is when multiple copies of the same sample are available, e.g., combinatorial spread libraries. In these, the local concentrations and functionalities are rigidly encoded in positional descriptor, allowing matching the regions between different tools. For these structures, only few characterization methods such as optical hyperspectral imaging or photoluminescence can be performed in parallel. For techniques such as structural characterization via focused X-Ray, chemical characterization via ToF-SIM, cathodoluminescence, scanning probe microscopy the measurements can be performed sequentially. At the same time, these techniques often give complementary information on structure, properties, and chemical composition. Correspondingly, running the automated experiment on same object and multiple systems allows to explore the material composition space combining information from multiple sources.

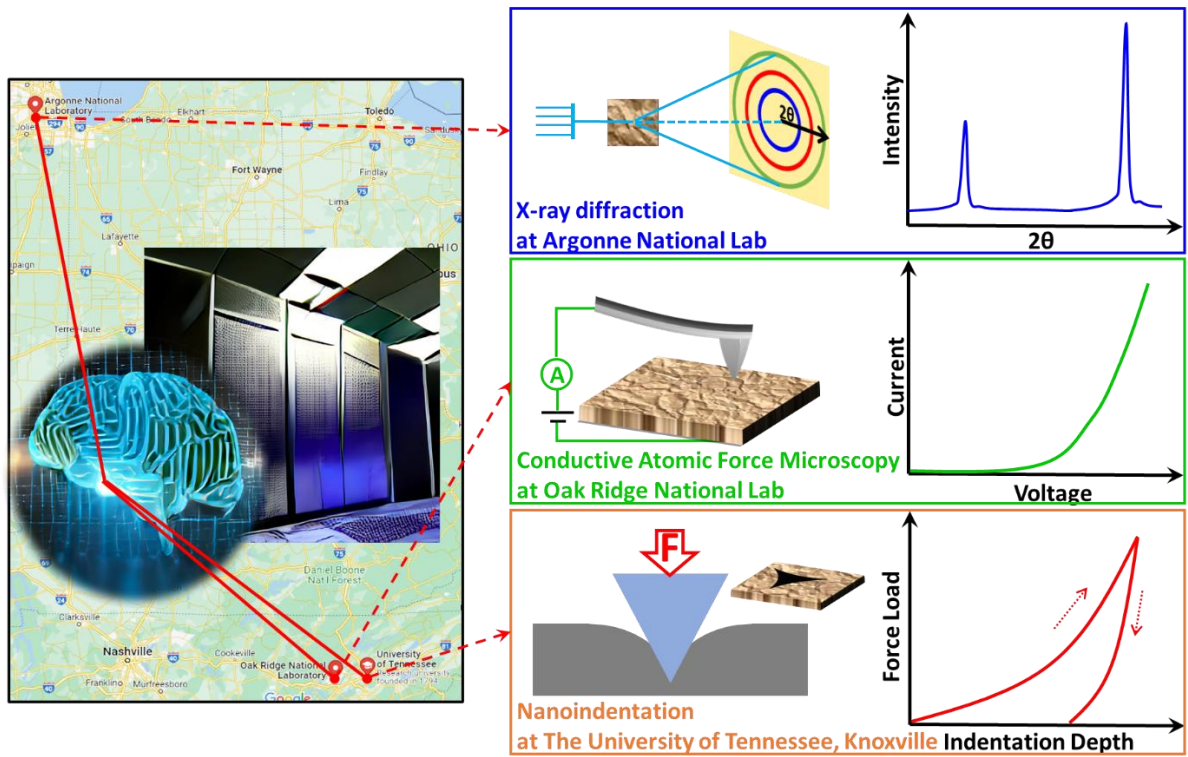

**Figure 8.** Graphic representation of the geographically distributed multimodal characterization experiment where the identical combinatorial library is simultaneously imaged via focused X-Ray at Argonne, nanoindentation at UT Knoxville, and SPM IV at ORNL and the experimental pathways are updated based on results from all three systems.

We pose that these experiments can be also performed in the statistical sense, in which the information from multiple tools is combined via partially similar channels. For example, for hybrid perovskite samples in Figure 1 the statistical properties of the microstructures can be explored via correlating the chemical or CL signals referenced to grain boundaries or other key point objects that can be identified in both methods.

**V.C. Automated laboratories**

The analysis above allows us to summarize the requirements for broad implementation of scientific workflow design for automated and traditional laboratories. These include:

1. Development of the labs capable of orchestrating predefined workflows based on human and robotic agents. These can be purely human operated lab, purely robotic lab, or human-robot lab where humans perform technical tasks. The associated need is the hyper language that summarizes possible operations and provides access to control parameters. This also includes process monitoring on multiple levels – sample locations, collection of proxy signals during processing, environment monitoring, etc.

2. Workflow design based on AI and human decision making, meaning specific series of synthesis and characterization steps based in hyperlanguage. Since physical objects that can be only in one place in one time, workflow will have a directed graph structure (but can form quasi loops when folded on material axis, e.g. for optimization). Note that currently humans both plan workflows and execute them, but these functions can be separated.
3. Defining domain-specific reward functions that guide workflow development. Why are we running experiments? Is the reward scientific discovery, optimization, or something else (curiosity or empowerment)? Ultimately, we should be able to quantify (in the style of Bellman equation for reinforcement learning) what is the benefit of the specific step in the workflow, and how does it accomplish or affects exploration and exploitation goals.
4. Integration of reward functions from dissimilar domains, since almost in all cases total reward function will be compounded from multiple intermediate rewards. For example, how does better microscope help us learn physics of specific material? Why would the specific DFT calculation help us understand experimental data?
5. Creating experimentally falsifiable hypothesis from the domain specific body of knowledge that can be incorporated in the exploratory part of automated workflows. This is required because workflow design should ideally include a discovery component, and not only optimization. Discovery is effectively extrapolation into unknown domain, and given very high dimensionality, the full space of possible experiments is intractable. Hypotheses provide a way to constrain the space to explore. Note that updating hypotheses based on experimental data is a well-defined problem in Bayesian sense.
6. Hypothesis generation. In many cases, this is an extrapolation problem, and will likely be human-driven (and potentially AI assisted) for the foreseeable future. However, it will be interesting to explore whether combination of interpolative capabilities of models such as ChatGPT combined with their ability to work with the full volume of text and other data available to mankind will be sufficient for hypothesis generation.

## VI. Summary

Scientific research and discovery are typically organized around workflows, or sequences of the specific actions and experiments targeting specific outcomes. Until now, the workflow design has been highly domain specific and once established, the workflows remain constant over decades. The disruption of the existing workflows or the introduction of the new ones is typically associated with the emergence of the novel experimental tools. Traditionally, ideation, orchestration, and implementation of the workflows is human based. The advent of machine learning tools over the last three years have facilitated optimization of human-built workflows, but yet have not led to beyond-human experimentation.

Here we introduce simple workflows for structural characterization and show that these can be based only on the discovery or weighted by prior knowledge. We suggest several possible strategies for the characterization workflow design. We discuss the increase of complexity for combined imaging-characterization workflows and illustrate the direct and inverse step for design of such workflow.

Finally, we argue that a similar approach can be extended to combined synthesis—imaging characterization workflows and workflows containing theory in the loop. The workflow design in this case becomes extremely complex, and will weigh the latencies, costs, and expected benefits of all steps. We believe that the design of such workflows will require careful analysis of the rewards and the analysis of the value of individual steps. However, the emergence of automated experiments and labs necessitates these developments. Overall, these tools will enable knowledge-based workflow optimization, enable lateral instrumental networks, sequential and parallel orchestration of characterization between dissimilar facilities, and empower distributed research.

**Acknowledgements**

This research (sections on mathematics of workflow optimization) was supported by the Center for Nanophase Materials Sciences (CNMS), which is a US Department of Energy, Office of Science User Facility at Oak Ridge National Laboratory. This effort was supported by the center for 3D Ferroelectric Microelectronics (3DFeM), an Energy Frontier Research Center funded by the U.S. Department of Energy (DOE), Office of Science, Basic Energy Sciences under Award Number DE-SC0021118. M.A. acknowledge support from National Science Foundation (NSF), Award Number No. 2043205 and Alfred P. Sloan Foundation, award No. FG-2022-18275. The authors acknowledge Peter Loxley, Tonio Buonassisi, Benji Maruyama, and Stephen R. Niezgoda for fruitful discussion.